\documentclass[runningheads]{llncs}
\DeclareUnicodeCharacter{2500}{\textemdash}
\usepackage[T1]{fontenc}
\usepackage{hyperref}
\usepackage{array}
\usepackage{graphicx}
\usepackage{adjustbox}
\usepackage{multirow}

\begin{document}
\title{Extracting Research Instruments from Educational Literature Using LLMs}
\titlerunning{Extracting Research Instruments from Educational Literature Using LLMs}
%
\author{Jiseung Yoo\inst{1}\and
Curran Mahowald\inst{2}\and
Meiyu Li\inst{3} \and
Wei Ai\inst{3}}
\authorrunning{Yoo et al.}
%

\institute{College of Education, University of Maryland, College Park, USA\\
\and
Annenberg Institute, Brown University, USA\\
\email{curran\_mahowald@brown.edu}
\and
College of Information, University of Maryland, College Park, USA\\
\email{\{jyoo20, ml0521, aiwei\}@umd.edu}}
\maketitle              

\begin{abstract}

Large Language Models (LLMs) are transforming information extraction from academic literature, offering new possibilities for knowledge management. This study presents an LLM-based system designed to extract detailed information about research instruments used in the education field, including their names, types, target respondents, measured constructs, and outcomes. Using multi-step prompting and a domain-specific data schema, it generates structured outputs optimized for educational research. Our evaluation shows that this system significantly outperforms other approaches, particularly in identifying instrument names and detailed information. This demonstrates the potential of LLM-powered information extraction in educational contexts, offering a systematic way to organize research instrument information. The ability to aggregate such information at scale enhances accessibility for researchers and education leaders, facilitating informed decision-making in educational research and policy.

\end{abstract}
\section{Introduction}

Identifying research instruments in the educational literature is crucial to synthesizing findings and ensuring replicability between studies. Research instruments are tools to collect, measure, and analyze empirical evidence corresponding with research purpose and questions~\cite{Wilkinson,Colton}. In education, effective measurement tools are essential for accurately capturing data on abstract concepts from student learning outcomes to school climate. These instruments allow researchers to collect standardized data across diverse populations and settings, enabling meaningful comparisons~\cite{Sturm}. In addition, well-documented instruments bridge research and practice, ensuring consistent use of evidence to inform decisions~\cite{Shaneyfelt}. For this reason, establishing a structured knowledge database to organize these instruments would enhance research interpretation, support consensus-building within the academic and practitioner communities, and help users adapt instruments to their specific contexts. It would also empower researchers and educators to make informed decisions when selecting measurement tools by providing clear, accessible knowledge about research instruments.

The identification and management of research instruments information in education remain underexplored, highlighting the need for systematic approaches. Databases like ERIC rely on manual expert annotation with predefined rules to categorize instruments by type, topic, and validity~\cite{ERIC,Cox}. Similarly, the Annenberg Institute’s EdInstruments system~\cite{EdInstrument} compiles existing research instrument data and integrates newly proposed tools from database users. However, given the continuous emergence of new instruments and the high cost of manual curation, a well-developed automated system is crucial. Traditional text analysis (e.g. n-grams) can be used for extraction but struggles with unstructured documents. Rule-based and machine-learning models rely on predefined patterns and hand-crafted lexicons, making them inflexible to text variations and requiring frequent manual updates ~\cite{Han,Gupta}. They also fail to capture deeper semantic connections, such as related concepts or implicit relationships. 

Recent advances in Large Language Models (LLMs) offer promising capabilities to address the issues, but naïve application often leads to inconsistent and unreliable results. Information extraction (IE) with the LLMs task involves automatically deriving structured information from unstructured text data, enabling machines to comprehend natural language text~\cite{Chuang}. The two main IE tasks are Named Entity Recognition (NER) and Relation Extraction (RE). NER identifies and classifies entities such as names and domain-specific terms, while RE detects and categorizes relationships between them. Although foundational LLMs and commercial platforms exhibit strong extraction and reasoning capabilities, they often struggle with hallucinations and domain specificity~\cite{Chen}. To mitigate these issues, studies have explored prompt engineering techniques and structured data schemas to enhance accuracy. Studies~\cite{Polak,Wei,Wu,Vatsal} show that specialized prompting—particularly iterative and multi-step methods—can significantly improve retrieval accuracy. Furthermore, studies ~\cite{Vijayan,Wang,Wiest,Chusova} demonstrate that constraining responses to predefined domain-specific schemas, such as JSON-based extraction, ensures structured and reliable output.

Building on these efforts, this study proposes a structured three-step prompt design with a domain-specific schema to systematically extract research instruments from educational papers. Our approach leverages education-specific schemas and an iterative prompt design to capture context-rich information. Also, by incorporating existing instrument information within the education context as a dictionary and detecting methods sections of research papers, our system accurately extracts instrument names and key relational details, including respondents, types, constructs, and outcomes. This work introduces a novel AI-assisted method for information extraction, enhancing accuracy and enabling large-scale analysis of educational research. The resulting knowledge repository provides researchers, schools, and district leaders with easy access to rich information on measurement tools, helping educators select appropriate instruments and generate high-quality evidence in their contexts.

\section{Method}

This study introduces a structured schema with multi-step, zero-shot prompting. The system identifies and links key entities, the instrument names, construct, outcomes, respondent, and instrument type, within their original contexts. 
This data schema enables users to interpret which tools were used, what they measured, and the target in each study. Research instruments fall into five categories: surveys/questionnaires, interview protocols, observation protocols, tests/assessments, and other tools (e.g. checklists)~\cite{Colton}. Each type includes essential components: constructs define the concept being studied, outcomes represent measured results, and respondents provide participant context. For instance, in anxiety research, constructs (e.g., "emotional distress"), outcomes (e.g., "anxiety level"), respondents (e.g., "undergraduate students"), and instrument type (e.g., "Likert-scale questionnaire") collectively define the measurement process.
\vspace{10pt}

\begin{figure}
\includegraphics[width=\textwidth]{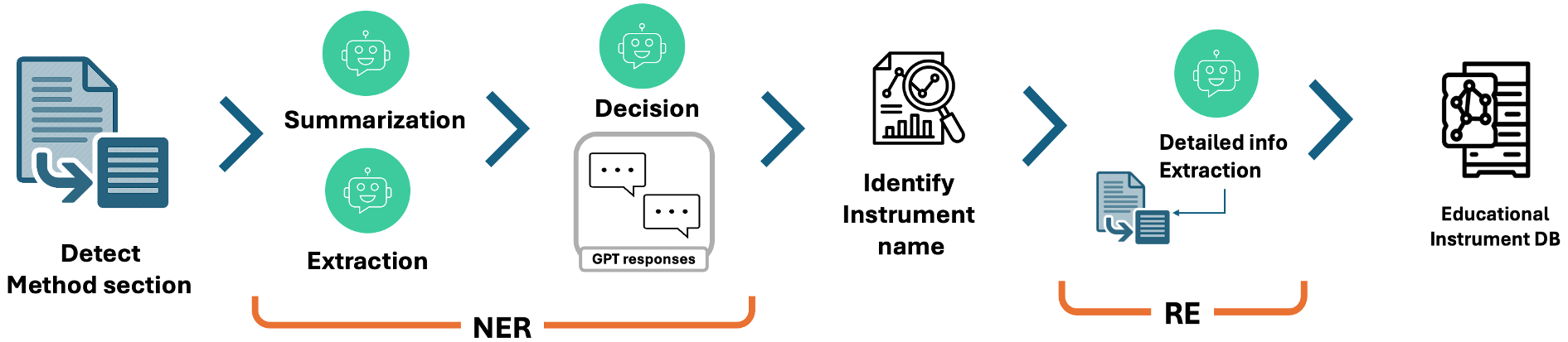}
\caption{Overview of the system pipeline.} \label{fig1}
\end{figure}
\subsection{Pipeline}

The system operates in a three-step pipeline. In the first step, it detects the methods section of research papers by using a pre-trained PDF parser model~\cite{Marker}, converting documents into a hierarchical JSON format. Since empirical social science research presents data collection and analysis in the methodology section, focusing on this section—rather than the full text—improves accuracy. The system identifies the start and end pages of the methods and results sections; if the methods section cannot be isolated, the full text is used. This detection process achieves 92\% accuracy (n=150), confirming its reliability. The extracted text is then split into 1000-token chunks to optimize processing efficiency with LLM APIs while preserving document structure. In the second step, the system employs multi-step prompting for NER to identify instrument names within the segmented text. Finally, in the third step, a RE prompt retrieves key details about the instruments with the extraction results being structured into a JSON format for downstream analysis. The next section will provide a detailed explanation of the second and third steps.

\subsection{Prompt Design}

This study uses iterative prompts and targeted follow-up questions to enhance context understanding and filter more accurate responses~\cite{Polak,Wei}. The NER process for instrument extraction employs a three-step prompting approach. First, an extraction prompt instructs the model to retrieve specific information about the instruments used in the study while providing background knowledge on research instrument concept (e.g. definition, purpose, and general usage). Next, a summarization prompt asks to explain how the study collects, measures, and analyzes data using these tools, ensuring a comprehensive understanding of their usage. Finally, a decision prompt consolidates and evaluates the outputs from the previous two prompts to generate a structured JSON output of key instruments. Once an instrument name is identified through NER, it is standardized using an instrument dictionary from Annenberg’s EdInstruments list before proceeding to the next stage.

For the subsequent relation extraction (RE) task, the identified instrument names serve as anchors, with the data schema embedded in the prompts to ensure consistent formatting and contextual accuracy. Using OpenAI function calling, the system extracts detailed information about each instrument, including its associated constructs, measured outcomes, instrument type, and target respondents.

\subsection{Evaluation}

Our experiment used an annotated dataset (n=150) from the Institute of Education Sciences (IES), U.S. Department of Education. This dataset contains manually labeled data, including instrument names and detailed information about instruments, making it a gold standard for evaluating entity extraction and relation linking. We used it to test various prompting strategies, including baseline approaches (ChatGPT-4o via web interface, zero-shot prompting), few-shot prompting, single-step prompts, and our proposed multi-step approach. These techniques were applied to both method section excerpts and full documents. Performance was assessed quantitatively and qualitatively, with three independent reviewers manually evaluating 30 randomly sampled documents for accuracy and error types. This evaluation allowed us to compare prompting methods and analyze trade-offs between accuracy, efficiency, and computational cost.

\section{Result}

\subsection{IE Performance}

Our system, using multi-step prompts with method section excerpts, performed competitively against few-shot prompts and general LLM outputs. Figure 2 shows NER performance across baselines, few-shot approaches, and various prompting strategies. The result suggests that targeting method excerpts can achieve strong performance while significantly reducing computational costs. On average, processing one document required 11,248 input tokens and 6,730 output tokens, 61\% fewer tokens than full-document processing, which used 26,181 input tokens and 20,109 output tokens. Additionally, method section excerpts with optimized prompting reduced processing time by 54.8\% compared to full-document processing.

\begin{figure}
\includegraphics[width=\textwidth]{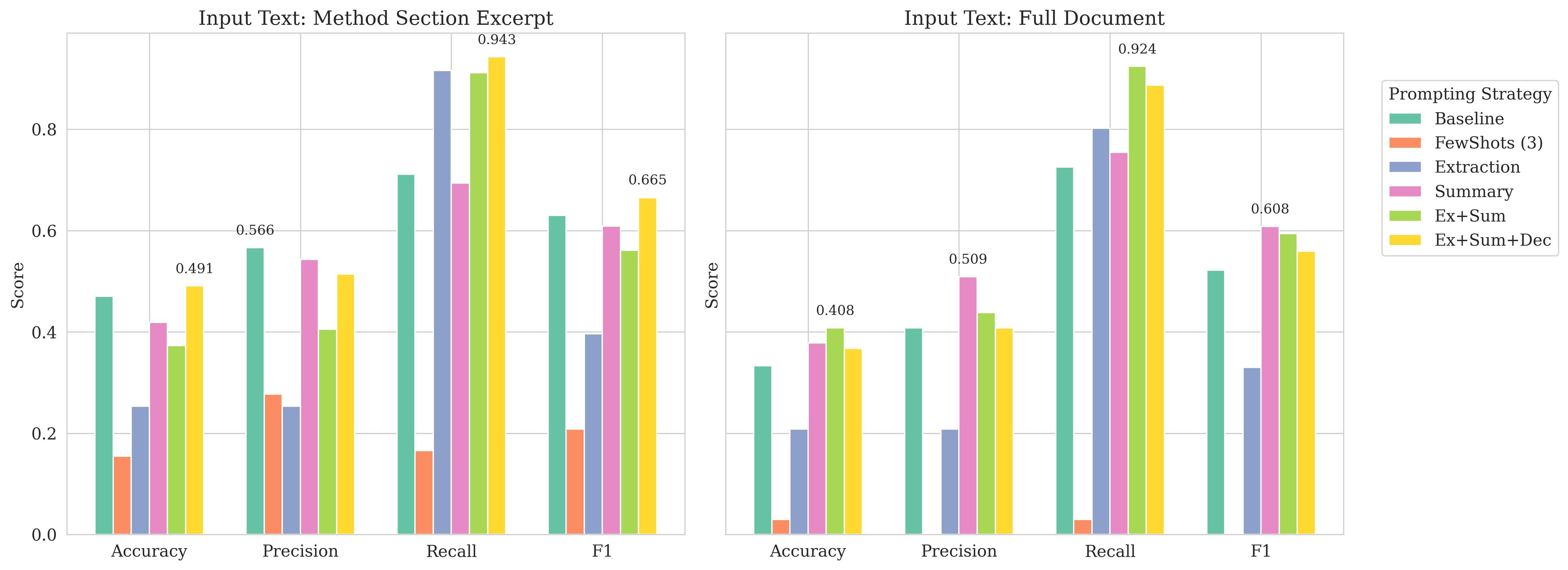}
\caption{Performance comparison across different prompts and input text types. `Ex' represents extraction, `Sum' represents summarization, and `Dec' represents decision. The highest F1 score (0.665) is achieved using a combination of summarization, extraction, and decision on the method section excerpt.} \label{fig2}
\end{figure}

Despite these efficiency gains, there is a tradeoff between recall and precision of the multi-step prompting. As shown in Table 1, GPT-o1 achieved the highest F1-score, with a recall of 90\%, effectively retrieving used instruments from research papers. However, its F1-score of 0.78 indicates lower precision, meaning that while the system identifies many instruments, some extractions are inaccurate or redundant.
\vspace{10pt}

\begin{table}[h!]
\centering
\caption{Performance of Different LLMs with Multi-Step Prompt and Method Excerpt}
\label{tab:performance}
\begin{tabular}{|>{\centering\arraybackslash}p{5cm}|>{\centering\arraybackslash}p{1.7cm}|>{\centering\arraybackslash}p{1.7cm}|>{\centering\arraybackslash}p{1.5cm}|>{\centering\arraybackslash}p{1.5cm}|}
\hline
\multicolumn{5}{|c|}{\textbf{Method Section Excerpt w/  Extraction + Summary + Decision}}\\ \hline
\textbf{Model}         & \textbf{Accuracy} & \textbf{Precision} & \textbf{Recall} & \textbf{F1}\\ \hline
Gpt-4o-mini   & 0.472  & 0.508  & 0.901  & 0.619  \\ \hline
Gpt-4o        & 0.491  & 0.514  & 0.943  & 0.665  \\ \hline
Gpt-o1        & 0.641  & 0.696  & 0.904  & \textbf{0.786}  \\ \hline
Claude-sonnet & 0.615  & 0.644  & 0.929  & 0.761  \\ \hline
Llama 3.3 70B & 0.396  & 0.608  & 0.639  & 0.623  \\ \hline
\end{tabular}
\end{table}

Table 2 highlights the system’s strengths and limitations in extracting structured information from educational literature. It accurately identifies the instrument name and type but sometimes broadens classifications, such as inferring "students" as respondents based on context. The system also expands constructs, capturing a broader set of related terms, whereas expert labels align with the CLASS framework’s predefined domains and dimensions. While the system's output enhances contextual understanding, it may reduce precision in distinguishing key constructs. For outcomes, the system identifies "teacher interaction" instead of the expert-labeled "classroom organization." However, given the original study’s focus on teacher-student interaction, the result remains relevant. Overall, the system effectively extracts and contextualizes instrument information.

\vspace{10pt}

\begin{table}[h!]
\centering
\caption{Example of Research Information Extraction Output}
\label{tab:performance_alt}
\renewcommand{\arraystretch}{1.2} 
\begin{tabular}{|>{\centering\arraybackslash}p{2cm}|>{\centering\arraybackslash}p{5cm}|>{\centering\arraybackslash}p{5cm}|}
    \hline
    \textbf{Category} & \textbf{Ground-truth} & \textbf{Model Output} \\ 
    \hline
    \textbf{Instrument} & CLASS (Classroom Assessment Scoring System) & CLASS (Classroom Assessment Scoring System) \\ 
    \hline
    \textbf{Type} & Observation Protocol & Observation Protocol \\ 
    \hline
    \textbf{Respondent} & Teacher & Students; Teachers \\ 
    \hline
    \textbf{Construct} & Behavioral Management & Classroom Organization, Preventive Discipline, Time Management \\ 
    \hline
    \textbf{Outcomes} & Classroom Organization & Teacher Interaction \\ 
    \hline
    
\end{tabular}
\end{table}

\subsection{Error Analysis}

Our error analysis revealed some challenges. First, the system's performance is influenced by the number and complexity of instruments mentioned in research papers. The dataset includes papers referencing an average of 3.66 instruments, with the system performing well when extracting 2 to 5. However, for papers labeled as having a single instrument, the system over-extracts, identifying an average of 6.5 instruments. In this case, the system often extracted sub-tests as independent instruments rather than recognizing them as part of a larger test battery. Second, the model’s sensitivity to context led to false positives, extracting instruments that were merely mentioned rather than actually used. Additionally, it prioritized information at the beginning of method sections, often overlooking instruments listed later. 

\section{Conclusion and Implication}

This work presents a structured pipeline for extracting research instruments from educational literature, showcasing how LLM-powered tools can address education-specific challenges. By automating instrument extraction, it facilitates large-scale synthesis of educational research, reducing manual effort and improving accessibility. A multi-step prompting approach with a domain-specific schema enhances precision, reliability, and interpretability over naive methods. Error analysis revealed challenges such as hierarchical misclassification and false positives, highlighting the need for refined ontological rules and human-in-the-loop validation. Despite these limitations, structured prompting significantly improves instrument identification. By making organized instrument data more accessible, this system helps researchers, educators, and policymakers select measurement tools suited to their purpose and contexts.

%
%
%
%

\end{document}